# Multiple symmetry protected BIC lines in two dimensional synthetic parameter space

Fengyuan Zhang[1], Qiongqiong Chu[1], Qiang Wang[1], Shining Zhu[1] and Hui Liu[1, *]

**Abstract:** Bound states in the continuum (BICs) have attracted significant interest in recent years due to their unique optical properties, such as infinite quality factor and wave localization. In order to improve the optical performance of BICs based devices, more degrees of freedom are required to tune BICs in high-dimension parameter space for practical applications. To effectively tune more BICs, we form a 2D synthetic parameter space based on a nanohole metasurface array. Multiple symmetry protected BIC modes with high Q factors can be achieved at high-order symmetry point. Through manipulating asymmetry parameters, BIC lines formed by a series of BIC modes can be found in the 2D synthetic parameter space. Moreover, the electric field distributions are investigated to demonstrate the generation and evolution of BICs. By measuring the absorption spectra, the tuning of multiple BICs with synthetic asymmetry parameters is experimentally explored, which agrees well with theoretical results. Therefore, our design can provide new insight for a variety of on-chip applications, such as non-linear devices, integrated nanolasing array and high-resolution sensors for infrared molecular detection.

**Keywords:** Bound states in the continuum, BIC lines, nanohole metasurface, synthetic parameter space

## 1 Introduction

BICs are eigenmodes embedded in the continuum radiation spectrum, generally characterized by infinite Q factor and perfect optical mode confinement [1-3]. This intriguing phenomenon was first proposed in quantum mechanics by von Neumann and Wigner [1], and has been found in many physical systems, such as electromagnetic, acoustic and water waves [4-6]. When small perturbations (such as structural parameters or incident angle) are introduced into the resonant system, symmetry protected BICs will transformation into quasi-BICs, exhibiting as high-Q resonances with tight optical confinement. The corresponding Q factors of leaky quasi-BIC modes depend on the asymmetry degrees caused by introduced perturbations [7].

In recent years, BICs have been extensively studied in various optical designs such as gratings [8-12], photonic crystals [13-16], waveguides [17-21], dielectric metasurfaces [22-25], and plasmonic systems [26-30], enabling numerous applications in nonlinear effect enhancement [31-32], lasers [33-35], sensors [36-37], and filters [9]. System with multiple BICs can provide more perspectives to flexibly manipulate the optical responses. New method for effectively tuning BICs in high-dimension parameter space is strongly needed for the development of various BIC based optical devices. Currently, the studies of optical mode modulation in high-dimension synthetic parameter space have become a hot topic due to the greatly increased tuning degrees of freedom. Based on 2D or 3D synthetic parameter space, our research group has achieved BIC mode modulation [38], charge-2 Dirac point in topological superlattices [39] and rotated Weyl physics probing [40-41].

On the basis of above researches, by introducing the concept of synthetic parameter space, which is formed by two independent asymmetry parameters, we demonstrate a system with multiple BICs in 2D synthetic parameter space through designed nanohole metasurface array. At high-order symmetry point, there are two BIC modes controlled by the x-direction asymmetry parameter and one BIC mode controlled by the y-direction asymmetry parameter. These BIC modes can achieve high Q factors which satisfies the common properties of symmetry-protected BICs. Away from this point, corresponding BIC mode will transform into quasi-BIC mode due to symmetry broken, manifesting as absorption peaks. Furthermore, BIC lines formed by a series of BIC modes can be achieved through adjusting the asymmetry parameters in proposed 2D synthetic parameter space. In experiments, measured absorption spectra under varied asymmetry parameters show the same trend as theoretical evolution of BICs. Benefiting from the continuously tunable optical responses of BICs, this proposed nanohole metasurface design can further enhance the light-matter interaction, opening a new avenue for plenty of physics fields in laser, sensing and nonlinear optics.

Fengyuan Zhang, Qiongqiong Chu, Qiang Wang and Shining Zhu, National Laboratory of Solid State Microstructures, School of Physics, Collaborative Innovation Center of Advanced Microstructures, Nanjing University, Nanjing, Jiangsu 210093, China.

# 2 Results and discussion

## 2.1 Multiple BICs in 1D parameter space

As shown in Figure 1a, our proposed multiple BICs system is composed of nanohole metasurface array, a gold mirror, and a dielectric Si layer sandwiched between them. In particular, each unit cell of artificially designed metasurface contains two identical nanoholes with fixed length L and width W as 2.2um and 0.2um, as shown in Figure 1c. For better understanding, we consider two nanoholes as two individual parts with the period along x (y) axis $P_x$ ($P_y$) fixed as $P_x = 1.5\ \mu m$ ($P_y = 3.6\ \mu m$). If we move one nanohole while the other keeps still, the symmetry of nanoholes will be broken. According to the moving distances $\Delta x$ and $\Delta y$, we define two asymmetry parameters $p = \Delta x / (P_x / 2)$ and $q = \Delta y / (P_y / 2)$ to form the 2D synthetic parameter space (p-q Space), as shown in Figure 1d. The SEM picture of one kind of asymmetry nanohole metasurfaces (p=0.4, q=0) is displayed in Figure 1b. In p-q space, the point of p=0, q=0 is the high-order symmetry point where the nanoholes simultaneously possess the $\sigma x$ and $\sigma y$ symmetry, which means that the structure is invariant under the mirror reflection $\sigma x$ and $\sigma y$, as illustrated in Figure 1c. With the increase of the parameters p and q, the electric field distribution of nanoholes changes accordingly to the increased asymmetry. When two nanoholes possess two out-of-phase electric dipole field distributions, they can be regarded as a nonradiative electric quadrupole. Specifically, when the electric field of two nanoholes manifests as two out-of-phase x-polarized (y-polarized) electric dipoles, we call it the x-polarized (y-polarized) quadrupole mode.

The mode properties of proposed multiple BICs system in 1D parameter space are firstly analyzed. We calculated the eigenfrequency variations of nanohole metasurfaces with varied asymmetry parameters p and q respectively, by COMSOL Multiphysics (COMSOL Inc.), as depicted in Figure 2a, b. In the simulation, the gold material is described by Drude model with plasma frequency set as $1.37 \times 10^{16}$ rad/s and the collision frequency $\Gamma$ set as 0 to realize lossless simulation. Here, we first focus on four modes at high-order symmetry point, two of which manifest as x-polarized electric quadrupole mode, that is, the BIC modes protected by $\sigma x$ symmetry, denoted as X1, X2. One mode manifests as y-polarized electric quadrupole, that is, the BIC mode protected by $\sigma y$ symmetry, denoted as Y1. The last ordinary mode is denoted as N1. Figure 2c shows the Ex distribution of four modes. The above three BIC modes and N1 mode are indicated by red and green circles, respectively.

To verify the existence of above three BIC modes, we further calculate the absorptivity of designed metasurfaces as a function of the asymmetry parameters p and q, respectively, as shown in Figure 2a, b. Here, the collision frequency $\Gamma$ of gold is set to $4.05 \times 10^{10}$ rad/s to obtain the absorption results, since the absorption of lossless system is nearly 0. It should be noted that this introduced metal loss is only one thousandth of the actual metal loss ($\Gamma_{Normal\ Au} = 4.05 \times 10^{13}$ rad/s), which is basically negligible. In Figure 2a, modes X1 and X2 have no corresponding absorption peaks at high-order symmetry point (indicated by the red circle), performing as nonradiative modes. However, when the parameter p changes continuously with q remains at 0, the absorption peaks corresponding to these two modes appear, performing as radiative modes. As parameter p moves away from the high-order symmetry point, the linewidth of absorption peaks gradually broaden, meaning that these two radiative modes become leakier. These results show that X1 and X2 are two nonradiative BIC modes at high-order symmetry point. Away from this point, X1 and X2 gradually evolve into quasi-BIC modes, manifesting as leaky modes due to broken $\sigma x$ symmetry. When p is fixed at 0 and q changes continuously, X1 and X2 modes have no corresponding absorption peaks and remain as BIC modes, as shown in Figure 2b. Clearly, these two BIC modes at high-order symmetry point are protected by $\sigma x$ symmetry (p=0). The slight discrepancies between the eigenfrequency and absorption peaks of modes X1 and X2 in Figure 2a are caused by the different $\Gamma$ set in the simulations. As shown in Figure 2c, the Ex of modes X1 and X2 at high-order symmetry point exhibit an odd symmetry distribution (x-polarized quadrupole), satisfying the central inverse symmetry (x, y) → (-x, -y). At this point, two nanoholes act as two radiation components that can cancel each other and then the radiation to the outside space is consequently forbidden. When $\sigma x$ symmetry is broken (p≠0), the symmetry of Ex distributions change from odd to even and the radiation channel to the outside space opens up.

Different from X1, X2 modes, the absorption peak of Y1 mode disappears at the high-order symmetry point and it will appear when q changes continuously while the parameter p is fixed as 0, as shown in Figure 2b. Therefore, this mode is a BIC mode at the high-order symmetry point and will evolve into quasi-BIC mode after moving away from that point. Similarly, when p changes continuously and q remains at 0, Y1 mode has no corresponding absorption peak and remains as BIC mode, indicating that this BIC mode is protected by $\sigma y$ symmetry (q=0), as shown in Figure 2a. In contrast, mode N1 always exhibits varying absorption peaks when p or q change continuously, which proves that this mode is a non-BIC mode (green circle), as shown in Figure 2a, b. The Ex of mode Y1 at high-order symmetry point similarly exhibits an odd symmetry distribution (y-polarized quadrupole) with no radiation to the outside space, as shown in Figure 2c. When $\sigma y$ symmetry is broken ($q\neq0$), the symmetry of Ex distribution changes from odd to even. However, the Ex of mode N1 always exhibits an even symmetry distribution, representing a radiative mode.

In order to reveal the underlying physical mechanism of these three BIC modes mentioned above, we calculated the variation of Q factors of them as the asymmetry parameters p or q change, as shown in Figure 2d, e. We can see that when q is fixed as 0 and p gradually varies, the Q factors of X1 and X2 modes reach the maximum values at the high-order symmetry point. After moving away from this point, these two modes transform into quasi-BIC modes and the Q factors of them drop rapidly. Moreover, the Q factors of two modes are stabilized near the maximum value when p is fixed as 0 and q gradually varies. In contrast, when p is fixed as 0 and q gradually varies, the Q factor of the Y1 mode reaches its maximum value at the high-order symmetry point and decreases rapidly after moving away from the point, indicating the evolution from BIC mode to quasi-BIC mode. Similarly, the Q factor of Y1 mode is stabilized near its maximum value when q is fixed as 0 and p gradually varies. For comparison, the N1 mode shows relatively low Q factor during the variation of parameters p or q. Then, through linear fitting, we find that the Q factors of X1, X2, and Y1 modes satisfy the relationship $Q \propto p^{-2}$ or $Q \propto q^{-2}$, further verifying that these three modes are symmetry-protected BIC modes at high-order symmetry point, as shown in Figure 2f, g. The Q factor of N1 mode do not satisfy the linear relationship with the negative quadratic of the asymmetry parameter p or q. The corresponding results for these modes of nanohole metasurfaces under considering metal loss are given in Figure S1.

## 2.2 BIC lines in 2D synthetic parameter space

Based on the above results in 1D parameter space, we continue to investigate the optical properties of nanohole metasurfaces in 2D parameter p-q space. Here, we call the symmetry points merely protected by the $\sigma x$ ($\sigma y$) symmetry where p=0, $q\neq0$ (q=0, $p\neq0$) as x-direction (y-direction) symmetry points. While the points where $p\neq0$, $q\neq0$, we call them as asymmetry points. Figure 3 shows the calculated Q factor variations of four modes X1, X2, Y1, and N1 in the p-q space, respectively.

In 2D parameter p-q space, it is clear that the maximum values of Q factors of X1 and X2 modes appear on the parameter line (p=0) connected by x-direction symmetry points. Away from this line, these two BIC modes transform into quasi-BIC modes, and the Q factors of them decrease continuously with the increase of |p|. Clearly, a BIC line is formed by a series of BIC modes which exist on the parameter line (p=0). Similarly, the maximum value of Q factor of Y1 mode appears on the parameter line connected by y-direction symmetry points. Away from this line, the Q factor of Y1 mode decreases continuously as |q| increases due to mode transformation. A new BIC line is formed by a series of BIC modes which exist on the parameter line (q=0). These BIC lines are indicated by red lines in the Figure 3.

The Q factor distribution in p-q space demonstrates that these BIC lines are protected by $\sigma x$ or $\sigma y$ symmetry. In addition, the Q factors of N1 mode present low values at all symmetry points, which is inconsistent with the BIC mode properties. In addition, we investigate the Q factors and the real part of the eigenfrequency variations of these four modes in p-q space with considering metal loss (as shown in Figure S2 and Figure S3). Along the BIC lines, BIC mode with highest Q factor can be rapidly found in p-q space, enabling the high-dimension tuning of optical performance. Benefiting from the increased tuning degrees of freedom provided by BIC lines, proposed nanohole metasurfaces can facilitate the developing of BIC based optical devices.

To better understand the physical mechanism of multiple BICs system, we select four points on the Q factor surface in p-q space for each of the four modes, namely, the high-order symmetry point (p=0, q=0), the x-direction symmetry point (p=0, q=-0.2), the y-direction symmetry point (p=-0.2, q=0) and the asymmetry point (p=-0.2, q=-0.2), for in-plane electric field distribution analysis, as shown in Figure 4. At high-order symmetry point, the Ex distribution of X1 and X2 modes in two nanoholes can be regarded as two out-of-phase electric dipoles, performing as x-polarized electric quadrupole. Similarly, the Ex distribution of Y1 mode performs as y-polarized electric quadrupoles. Those non-radiative quadrupole distributions demonstrate the simultaneous existence of three BIC modes. At x-direction symmetry point, only the Ex distributions of X1, X2 modes are maintained as electric quadrupoles while the Ex distribution of Y1 mode in two nanoholes is in-phase, indicating the transformation from BIC mode to quasi-BIC mode. On the contrary, at y-direction symmetry point, only Y1 mode maintains as electric quadrupole while the X1, X2 modes are no longer electric quadrupole distribution. However, at asymmetry point where the symmetry is broken in both directions, the Ex distributions of X1, X2 and Y1 modes exhibit same in-phase properties and all transform into radiative quasi-BIC mode. Differently, the Ex distribution of N1 mode at four picked points exhibits the same in-phase distribution, performing as radiative non-BIC mode. To clearly show the radiation mechanism of BIC modes, we also provide the out-of-plane (x-z plane) electric field Ex and |E| distributions for each of the four modes at four points in p-q space, shown as Figure S4 and Figure S5.

## 2.3 Experiment results and analysis

In the following, we introduced metal loss into nanohole metasurfaces and simulated the absorption spectra of metasurfaces under varied asymmetry parameters to verify the existence of BIC lines in lossy system. The absorption spectra under varied asymmetry parameter p (q=0) were firstly simulated, as shown in Figure 5a. It can be seen that there is no resonant absorption peak for the X1 and X2 modes at the high-order symmetry point. After moving away from this point, the absorption peaks of X1, X2 modes appear and show redshift and blue shift respectively, indicating the evolution from BICs to quasi-BICs. The absorption peaks of two modes are marked by the red dots in the figure. In contrast, the absorption peaks of N1 mode always present and its resonance wavelength remains almost unchanged. Above results are consistent with that in lossless system in Figure 2a. Note that, compared to lossless system, the absorption peaks of lossy system are broadened due to the presence of metal loss.

Specifically, we fabricated designed nanohole metasurface array with varied asymmetry parameter p and measured the absorption spectra, as shown in Figure 5b. The variation tendency of absorption peaks of X1, X2 modes show good agreement with the simulation results, verifying the existence of two BIC modes in experiments. At the same time, the Y1 mode shows no corresponding absorption peak and thus maintain as BIC mode, which means that there is a BIC line formed by a series of BICs under varied p. Moreover, the Q factors of X1, X2 modes in experimental and simulated results are compared, as shown in Figure 5c, d. Both of them show the same trend and are consistent with the properties of the $\sigma x$ symmetry protected BIC modes. The slight discrepancies between the Q factors of simulation and experimental results are caused by fabrication errors.

Then, we investigate the absorption spectra variations under varied asymmetry parameter q (p = 0), as shown in Figure 6a, b. It can be seen that the experiment results agree well with the simulation results. There is no absorption peak for the Y1 mode at high-order symmetry point, but away from this point, corresponding absorption peaks appear and gradually shift, indicating the appearance of quasi-BIC. The variation tendency is same as that of lossless system in Figure 2b. Meanwhile, the X1, X2 modes show no corresponding absorption peak and thus both maintain as BIC modes, which means that there are BIC lines formed by a series of BIC modes under varied q. In Figure 6c, the Q factors of experiment results show the same tendency as that of simulation results and are consistent with the properties of the $\sigma y$ symmetry protected BIC modes.

Above results clearly show the generation and evolution of BIC lines in p-q space at the condition of considering metal loss. Therefore, our proposed multiple BICs system provides more degrees of freedom for BICs tuning, offering a new platform for multifunctional integrated optical devices. For example, by introducing this design method into semiconductor nanolasing regime, low-threshold integrated nanolasing array are expected to be realized. In addition, most studied BIC-based sensors have focused on the wavelength range from visible to near-infrared. However, our design can basically cover the long-wave infrared range of 8-14um by asymmetry parameter manipulation, which can be expected to be applied for integrated high-resolution sensors for infrared molecular detection.

## 3 Conclusion

In summary, based on nanohole metasurface array, we have realized multiple BIC lines formed by a series of BICs in 2D synthetic parameter space. Desired BIC mode with highest Q factor can be found along the BIC lines. Through adjusting the asymmetry parameters in p-q space, the continuously tuning of BICs are simultaneously realized in simulation and experiments. Specifically, the physical mechanisms underlying the generation and evolution of BICs have been carefully investigated. This proposed system with multiple BICs can find many potential optical devices in nonlinear optics, nanolasing and infrared sensing. Meanwhile, our proposed new method for effectively tuning BICs in high-dimension synthetic parameter space can be applied to various BIC systems, paving the way for multi-dimensional manipulation.

## 4 Methods

### 4.1 Nanohole metasurface array fabrication

Firstly, three layers of Au (70 nm)/Si(300 nm)/Au(100 nm) are sequentially deposited on the Si substrate by Electron Beam Evaporation (AdNaNotek EBS-150U). Then nanohole metasurface are etched from the top gold film by a focused ion beam (FIB dual-beam FEI Helios 600i, 30 keV, 100 pA). By adjusting the etching parameters, the asymmetric parameters p and q of nanohole metasurface array can be varied from -1 to 1 while the etching depth is set as 70 nm. Each sample of the metasurface array in p-q space is 100 μm×100 μm sized.

## 4.2 Optical characterization

The absorption spectra (A) are derived based on the measured reflection spectra (R) by A = 1 − R. Specifically, each reflection spectrum (R) of the nanohole metasurface array was measured by a Fourier transform infrared (FTIR) spectrometer. The reflection signals in the spectral range of 4− 16 μm were collected using a Hyperion 2000 IR microscope with a liquid-nitrogen-cooled HgCdTe (MCT) detector. Measured reflection spectra were normalized with respect to a gold mirror.

## 4.3 Numerical Simulations and Analysis.

To find eigenmodes of multiple BICs system, numerical simulations are performed to calculate the eigenfrequency of nanohole metasurface array by the eigenfrequency solver in COMSOL Multiphysics. The Q factor is obtained from the real and imaginary parts of the eigenfrequency:

$$Q = \frac{real\ (Eigenfrequency)}{2 \times imag\ (Eigenfrequency)}$$

The absorption of each individual metasurface can also be simulated. During the process, periodic boundary conditions are applied in both x and y directions for mimicking the periodic nanohole array.

**Author contribution:** All the authors have accepted responsibility for the entire content of this submitted manuscript and approved submission.

**Research funding:** This work was financially supported by the National Natural Science Foundation of China (Nos. 92163216 and 92150302).

**Conflict of interest statement:** The authors declare no conflicts of interest regarding this article.

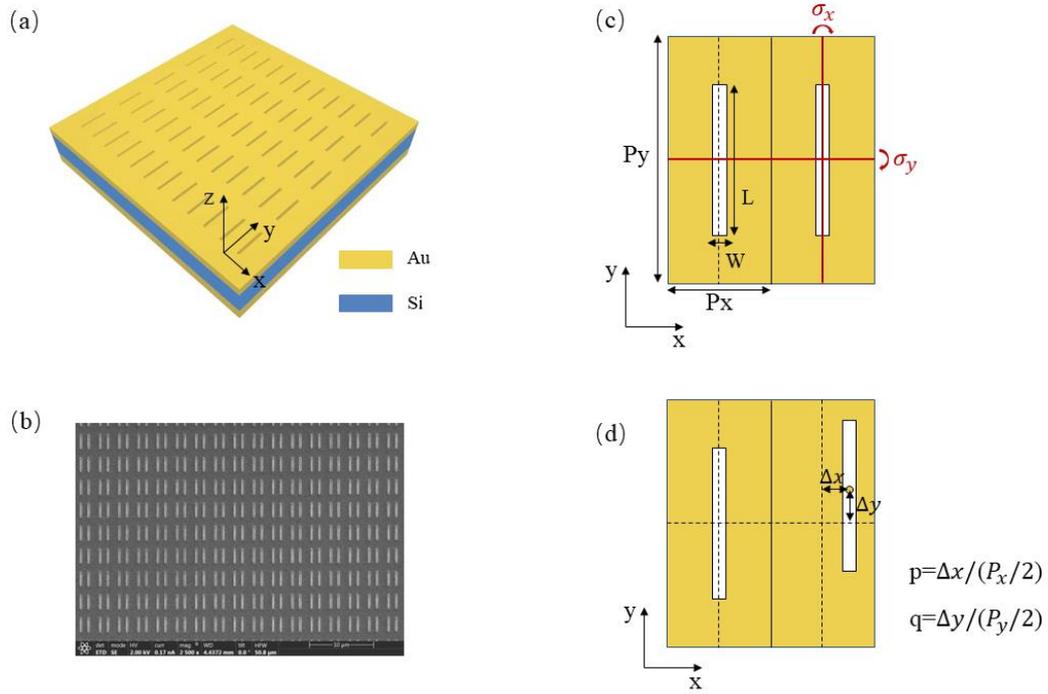

Figure 1. Schematic diagram of multiple BICs system based on nanohole metasurface array. (a) Schematic of designed nanohole metasurface. (b) The SEM picture of fabricated one kind of asymmetric nanohole metasurfaces. (p=0.4, q=0) The unit cell of nanohole metasurface at high-order symmetry point where nanoholes simultaneously possess $\sigma_x$ and $\sigma_y$ symmetry (c) and the asymmetry points where the symmetry of nanoholes is broken (d).

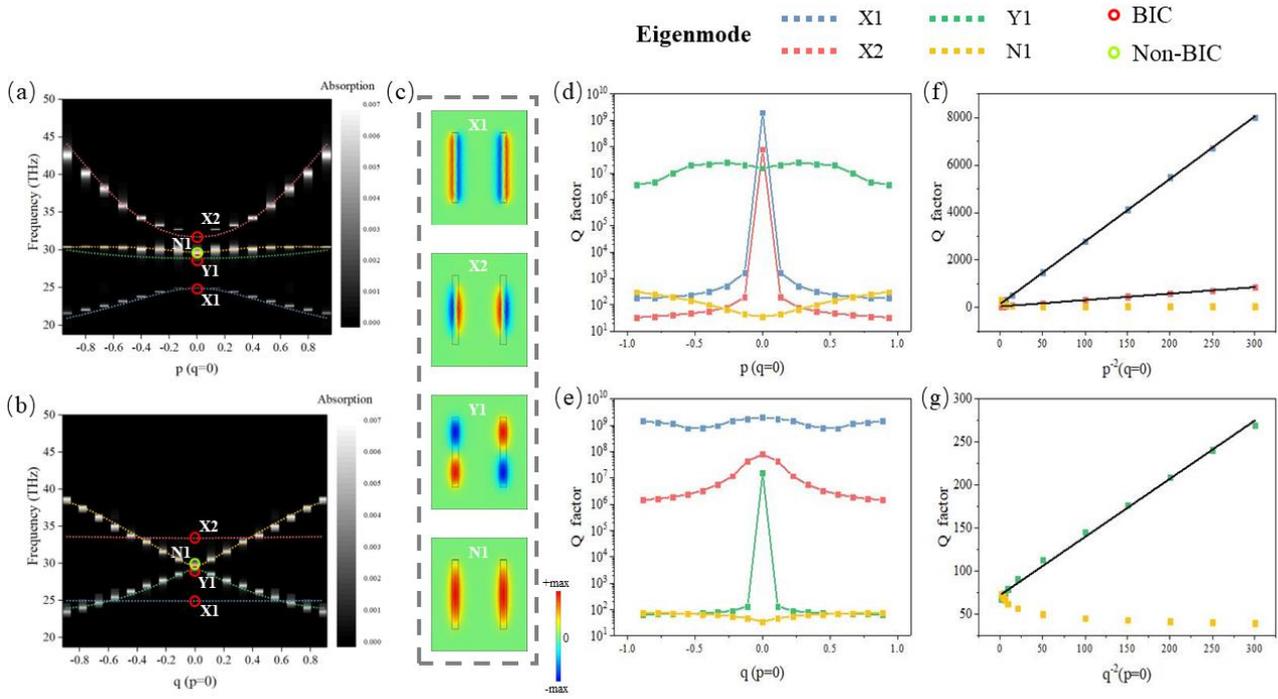

Figure 2. Multiple BIC modes investigation in 1D parameter space. (a) Simulated eigenfrequency variations of four modes X1, X2, Y1, N1 and corresponding absorption spectra variations with varied asymmetry parameter p while q fixed as 0. The position of the BICs (non-BIC) are marked by red (green) circles. (b) Simulated eigenfrequency variations of four modes and corresponding absorption spectra variations with varied asymmetry parameter q while p fixed as 0. (c) The Ex distribution of four modes at high-order symmetry point (p=0, q=0). Variations of Q factors with varied asymmetry parameter p (q=0) (d) and asymmetry parameter q (p=0) (e) respectively for four modes. (f) Linear fit of the Q factor of modes X1 and X2 to the $p^{-2}$. (g) Linear fit of the Q factor of mode Y1 to the $q^{-2}$.

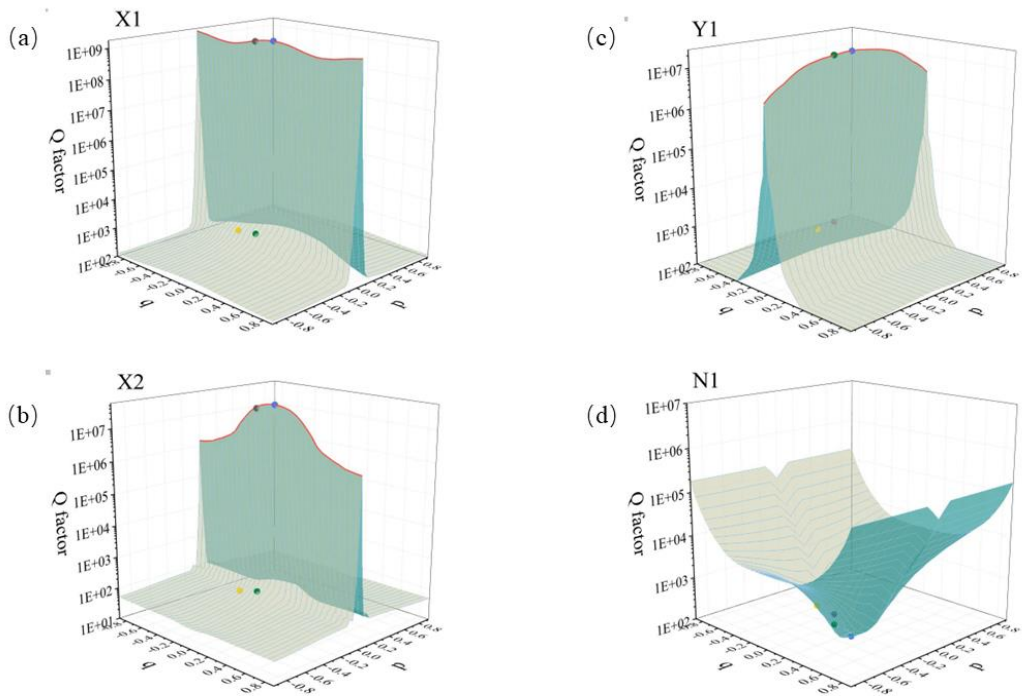

Figure 3. Multiple BIC modes investigation in synthetic p-q space. Distribution of Q factors in p-q space for (a) X1, (b) X2, (c) Y1 and (d) N1 modes. BIC lines are marked by red lines. Four picked points for field analysis are marked by colored dots.

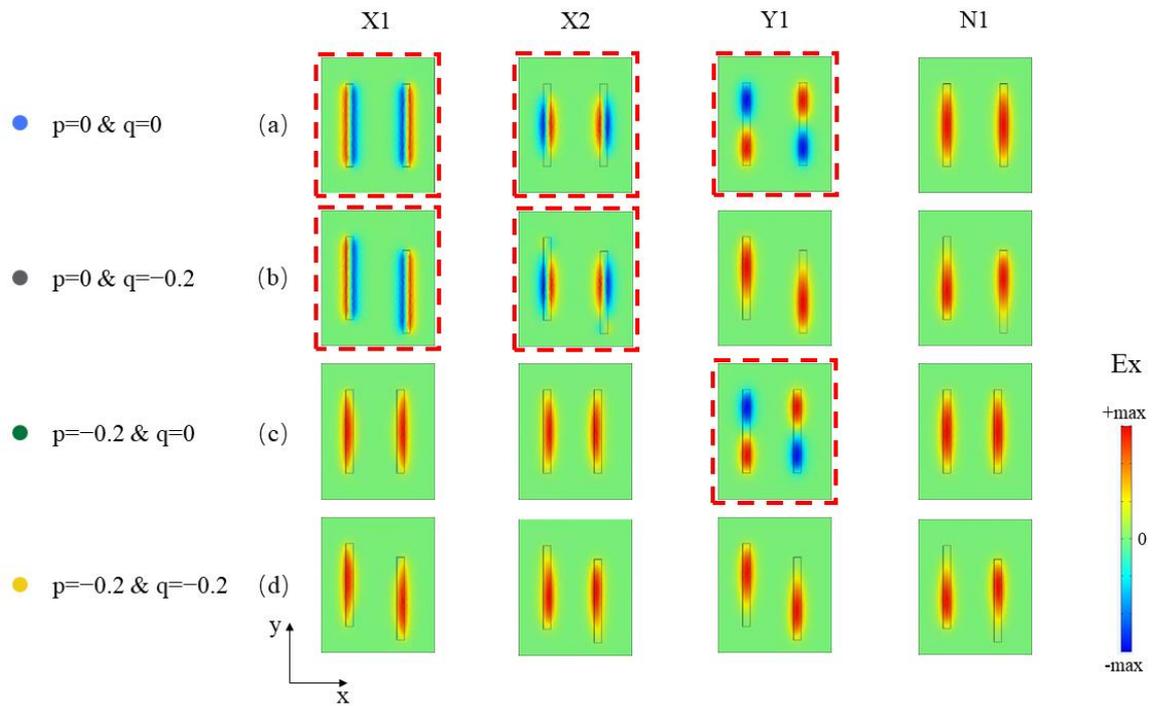

Figure 4. Electric field distribution of modes in p-q space. The in-plane Ex distribution for X1, X2, Y1 and N1 modes at four picked points where p=0, q=0 (a), p=0, q=-0.2 (b), p=-0.2, q=0 (c) and p=-0.2, q=-0.2 (d). The red dashed box represents the BIC modes.

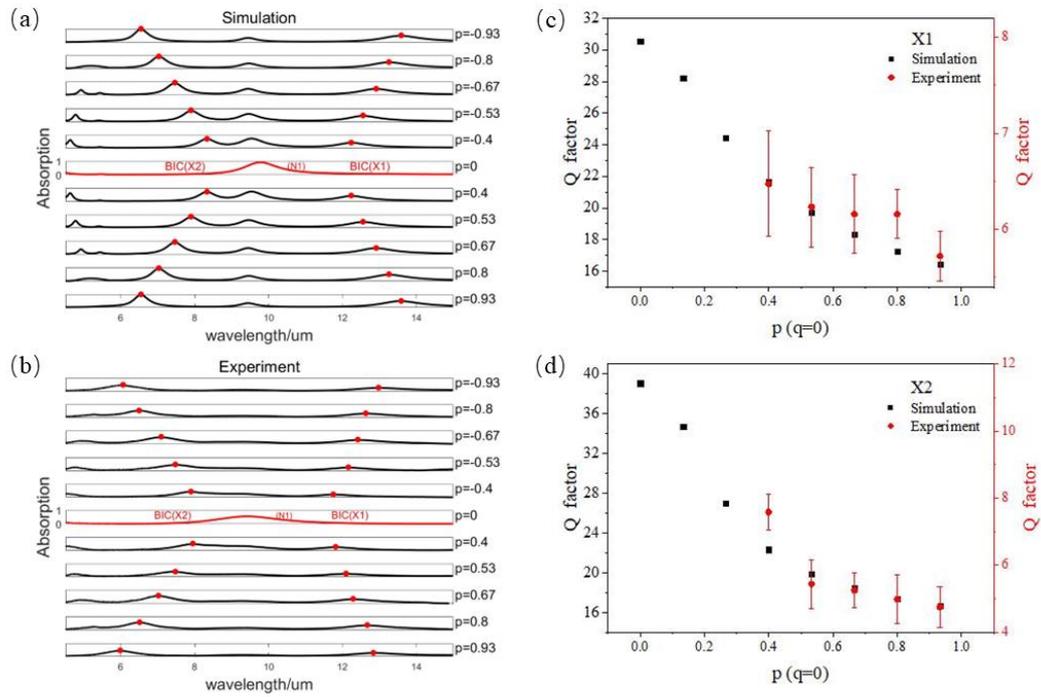

Figure 5. Experimentally demonstrated multiple BICs system in p-q space. Simulated absorption spectra in lossy system (a) and experimentally measured results (b) of nanohole metasurfaces under varied asymmetry parameter p (q=0). The red line indicates the nonradiative BIC modes (X1 X2) and the red dots indicate the absorption peaks corresponding to the quasi-BIC modes. Comparison of Q factors obtained from simulation and experiment for X1 (c) and X2 (d) modes.

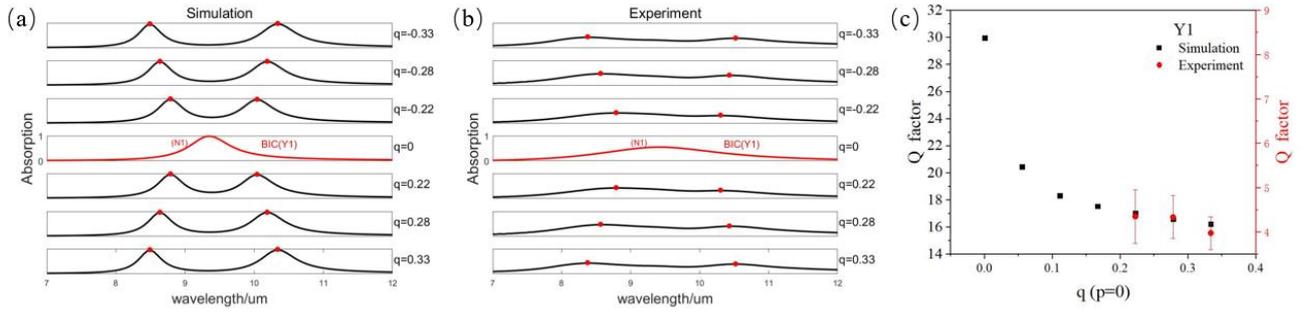

Figure 6. Experimentally demonstrated multiple BICs system in p-q space. Simulated absorption spectra in lossy system (a) and experimentally measured results (b) of nanohole metasurfaces under varied asymmetry parameter q (p=0). (c) Comparison of Q factors obtained from simulation and experiment for Y1 mode.

## Supplementary Material

To demonstrate the existence of multiple BIC modes and their properties in the lossy regime, we first calculated the eigenfrequency variations of nanohole metasurfaces with varied asymmetric parameters p and q respectively with considering metal loss, as depicted in Figure S1a, b. The variation of the eigenfrequencies of the four modes is consistent with that in the lossless regime. Figure S1c shows the Ex distribution of four modes. Three BIC modes X1, X2, Y1 and N1 mode are indicated by red and green circles, respectively.

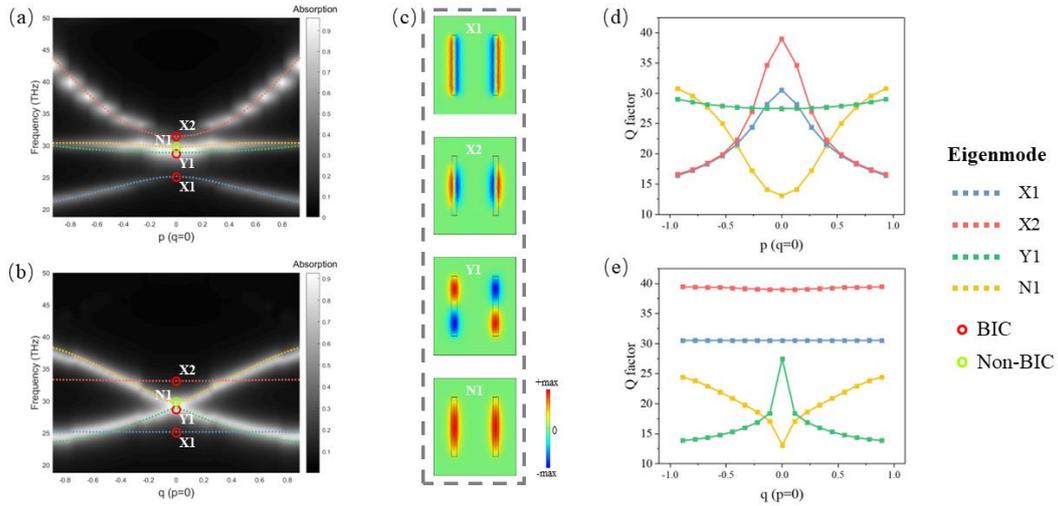

Figure S1. Multiple BIC modes investigation in 1D parameter space with considering the metal loss. (a) Simulated eigenfrequency variations of four modes X1, X2, Y1, N1 and corresponding absorption spectra variations with varied asymmetry parameter p while q fixed as 0. The position of the BICs (non-BIC) are marked by red (green) circles. (b) Simulated eigenfrequency variations of four modes and corresponding absorption spectra variations with varied asymmetry parameter q while p fixed as 0. (c) The Ex distribution of four modes at high-order symmetry point (p=0, q=0). Variations of Q factors with varied asymmetry parameter p (q=0) (d) and asymmetry parameter q (p=0) (e) respectively for four modes.

Then, we calculate the absorption spectra of designed metasurfaces as a function of the asymmetric parameters p and q, respectively, as shown in Figure S1a, b. The absorption spectra of each mode in lossy system show the same variation tendency as that in the lossless system, except for the broaden linewidth of absorption peaks due to metal loss. we also calculated the Q factor variations of these modes as the asymmetric parameters p or q change, as shown in Figure S1d, e. We can see that the results are also consistent with that in the lossless system, except for the decreased Q factors due to metal loss.

The calculated Q factors and real part of eigenfrequency of four modes X1, X2, Y1, and N1 in p-q space with considering the metal loss are shown in Figure S2 and Figure S3, respectively. Due to the metal loss, the Q factors of BICs are decreased but vary with same tendency as that in the lossless system. As for the real part of eigenfrequency, with varied asymmetry parameters p and q, four modes exhibit different characteristics. Thus, by modulating the asymmetry parameters in p-q space, both in the lossless and lossy system, proposed nanohole metasurface array design can achieve flexibly optical properties tuning of BICs.

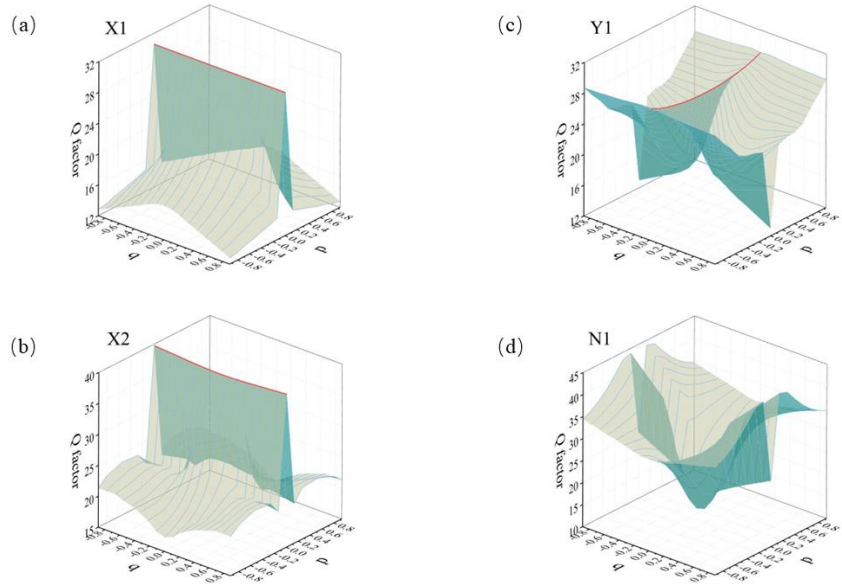

Figure S2. Multiple BIC modes investigation in synthetic p-q space with considering the metal loss. Distribution of Q factor in p-q space for (a) X1, (b) X2, (c) Y1 and (d) N1 modes. BIC lines are marked by red lines.

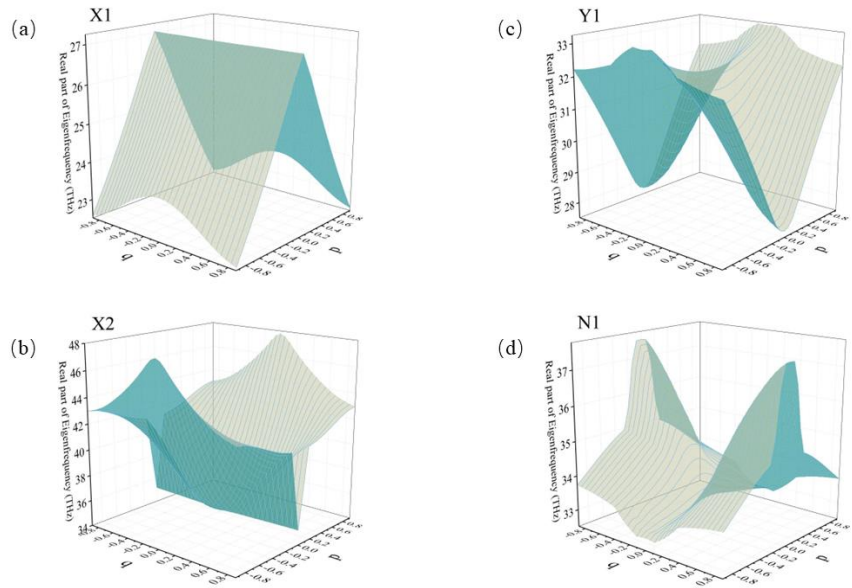

Figure S3. Multiple BIC modes investigation in synthetic p-q space with considering the metal loss. Distribution of the real part of eigen-frequency in p-q space for (a) X1, (b) X2, (c) Y1 and (d) N1 modes.

For further verify these modes (X1, X2 and Y1) under symmetry protection are BICs, we analyzed the out-of-plane electric field distribution of the discussed modes. The electric field Ex and |E| distributions in the x-z plane (y=0 cross section) for each of the four modes at four points in p-q space, are shown in Figure S4 and Figure S5, respectively. The BIC modes are marked by red dashed box. It can be seen that the field distributions of these BIC modes are all localized inside the structure with no outward radiation. When p or q changes, the electric field distribution appears outside the structure, which means that the quasi-BIC modes start to radiate outward. In contrast, for non-BIC mode N1, outward radiation always exists at all points. Therefore, the above results can prove that modes X1, X2 and Y1 under symmetry protection are BICs.

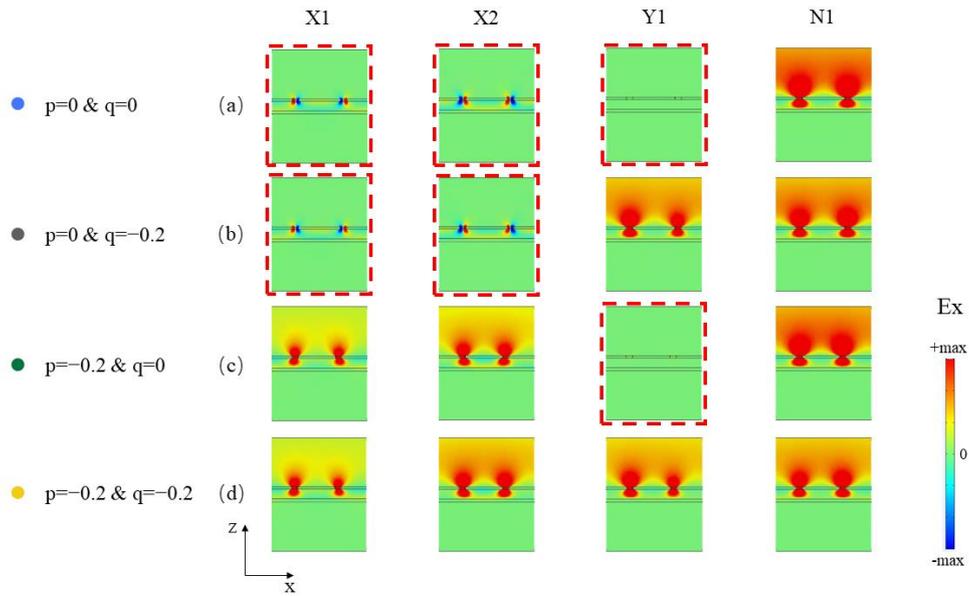

Figure S4. Electric field distribution of four modes in p-q space. The out-of-plane Ex distribution for X1, X2, Y1 and N1 modes at four picked points where p=0, q=0 (a), p=0, q=-0.2 (b), p=-0.2, q=0 (c) and p=-0.2, q=-0.2 (d). The red dashed box represents the BIC modes.

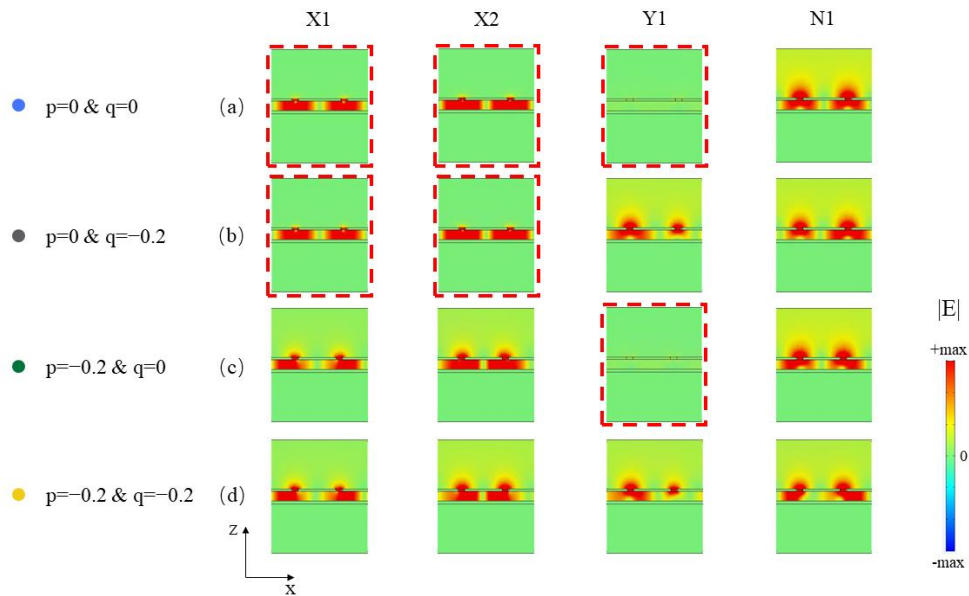

Figure S5. Electric field distribution of four modes in p-q space. The out-of-plane |E| distribution for X1, X2, Y1 and N1 modes at four picked points where p=0, q=0 (a), p=0, q=-0.2 (b), p=-0.2, q=0 (c) and p=-0.2, q=-0.2 (d). The red dashed box represents the BIC modes.